\documentclass[useAMS,usenatbib,usedcolumn,usegraphicx]{mn2e}
\usepackage{xspace}
\usepackage{color,epsfig,rotating,amssymb,longtable}
\usepackage{array,colortbl}
\usepackage[colorlinks=true,linkcolor=magenta,citecolor=blue,breaklinks=true]{hyperref}

\usepackage{ifpdf}

\ifpdf
\usepackage{epstopdf}
\else
\usepackage[dvips]{graphicx}
\usepackage{lscape}
\fi

\newcommand{\dd}{{\rm d}}

\newcommand{\HII}{H{\sc ii}\xspace}
\newcommand{\Ha}{\ensuremath{{\rm H\alpha}}\xspace}
\newcommand{\Msun}{\ensuremath{{\rm M_\odot}}\xspace}

\newcommand{\Ne}{\ensuremath{n_{\rm e}}}

\usepackage{color}

\definecolor{dgreen}{rgb}{0,.5,.1} 
\definecolor{pink}{rgb}{.9,.2,.5}  
\definecolor{orange}{rgb}{.9,.4,0} 
\definecolor{darkred}{rgb}{.545,0.0,.0}
\title[Sub-arcsecond mapping in NGC\,3351]
{Sub-arcsecond radio continuum mapping in and around the spiral galaxy
  NGC\,3351 using MERLIN}

\author[G. F. H\"agele et al.]
{Guillermo F. H\"agele$^{1,2}$\thanks{E-mail: guille.hagele@uam.es}, Yago
 Ascasibar$^{1}$, Anita M.\ S.\ Richards$^{3}$, 
\newauthor M\'onica V. Cardaci$^{1,4,2}$,
 Javier V\'asquez$^{5,2}$, \'Angeles I. D\'{\i}az$^{1}$,
\newauthor Daniel Rosa Gonz\'alez$^{6}$, Roberto
Terlevich$^{6}$\thanks{Research Affiliate, IoA, University of Cambridge, UK}
and Elena Terlevich$^{6}$\thanks{Visiting astronomer at IoA} 
\\
$^{1}$ Departamento de F\'{\i}sica Te\'orica, Universidad Aut\'onoma de
Madrid, 28049 Madrid, Spain\\ 
$^{2}$ Facultad de Cs.\ Astron\'omicas y Geof\'isicas, Universidad Nacional de La
Plata, Paseo del Bosque s/n, 1900 La Plata, Argentina \\ 
$^{3}$ Jodrell Bank Centre for Astrophysics, School of Physics and Astronomy,
University of Manchester M13 9PL UK\\
$^{4}$ XMM Science Operations Centre, European Space Astronomy Centre of ESA,
P.O. Box 50727, 28080 Madrid, Spain\\ 
$^{5}$ Instituto Argentino de Radioastronom\'ia (CCT-La Plata, CONICET),
C.C.5., 1894 Villa Elisa, Argentina\\
$^{6}$ Instituto Nacional de Astrof\'isica, \'Optica y Electr\'onica,
Tonantzintla, Apdo. Postal 51, 72000 Puebla, M\'exico\\ } 

\begin{document}

\maketitle

\begin{abstract}

We report sub-arcsecond scale radio continuum observations of a field of 35
by 22\,arcmin centred in NGC\,3351 obtained with the Multi-Element Radio
Linked Interferometer Network (MERLIN). We found 23 radio sources in this
field, 6 of which are projected within the D$_{25}$ isophote of the galaxy,
and 3 are located inside the central 100\,arcsec in radius. 
Two of these three are significantly extended, while the third one is relatively
compact. This one is the only source with a previously detected counterpart at other
wavelengths and could constitute the radio counterpart of a young supernova
remnant. The other two are probably related to jets from a
background AGN. We are not able to detect individual supernovae or SNRs in the
central region ($r<600$ pc) of the galaxy. This could imply that the ionising
populations of the circumnuclear
star-forming regions (CNSFRs) are too young (less than a few Myr) to host
supernovae. Also the diffusion length of the relativistic electrons in the ISM
associated with the SN from previous events of star formation 
seems to be larger than our maximum resolution of 50 pc in radius.
Detecting the thermal bremsstrahlung emission from the circumnuclear \HII
regions probably requires deeper observations.

\end{abstract}


\section{Introduction}
\label{secIntro}

The radio emission of a normal spiral galaxy consists of a thermal
bremsstrahlung component, arising from inelastic Coulomb collisions between
the free electrons and ions of the interstellar medium (ISM), and a
synchrotron component emitted by relativistic cosmic-ray electrons from supernova
remnants (SNRs) and active galactic nuclei (AGN) spiralling
in the galactic magnetic field \citep[see e.g.][]{Condon92}.
Star-forming regions may host a large number of supernovae, and the
\HII regions created by young massive stars can be detected as thermal sources
\citep[e.g.][]{KobulnickyJohnson99,Turner+00,Johnson+01,Johnson+03,Johnson+09,CannonSkillman04,Tsai+06,Tsai+09,Rosa-Gonzalez+07a}. 
In particular, the inner ($\sim1$~kpc) parts of some spiral galaxies show
higher star formation rates (SFRs) than usual, frequently arranged in a ring pattern
around the nucleus. Circumnuclear star-forming regions (CNSFRs), also referred
to as ``hot-spots'' \citep{Sersic+65,Sersic+67}, are similar to
luminous and large disk \HII regions, but they are more compact and show
higher peak 
surface brightness \citep{Kennicutt+89}. Their massive stars can dominate the
observed ultraviolet (UV) emission even in the presence of AGN
\citep{Gonzalez-Delgado+98,Colina+02}, and the ionised gas can contribute up
to 50 per cent to the total H$\beta$ luminosity in some Seyfert~2 nuclei
\citep{Cid-Fernandes+01}. The \Ha luminosities, typically higher than
$10^{39}$~erg~s$^{-1}$, overlap with those of \HII galaxies
\citep[see e.g.][and references therein]{Melnick+88,Diaz+00,HoyosDiaz06}. Due
to their location near the centre of the galaxy, CNSFRs display high metal
abundance \citep[see e.g.\ ][and references therein]{Diaz+07}, which makes
them ideal laboratories to investigate star formation in environments of large
metallicity.

NGC\,3351, also known as M95 and UGC5850, is a nearby
\citep[10.05\,Mpc, 49 pc/arcsec,][]{Graham+07} SBb(r)II early-type barred spiral galaxy
\citep{Sandage+87} in the Leo group. 
Its coordinates are $\alpha_{\rm 2000}=10^{\rm h}\,43^{\rm m}\,57\fs7$, $\delta_{\rm
2000}=+11^{\circ}\,42\arcmin\,12\farcs7$, and its major- and minor-axis
diameters at the 25 magnitude isophote (D$_{25}$) are
$7.4 \times 5.0$~arcmin, with a position angle (PA) of 13$^{\circ}$
\citep[RC3 catalogue,][]{deVaucouleurs+91}. NGC\,3351 has a star-forming  
circumnuclear ring, and has been classified as a ``hot-spot'' galaxy by
\cite{Sersic+67}. Early detailed studies of its nuclear regions by
\cite{Alloin+82} showed that NGC\,3351 harbours high-mass star
formation along a ring of about 20\,arcsec in diameter in its central zone.
In fact, it can be considered a nuclear starburst galaxy, since the SFR per unit
area is significantly higher in the nuclear region than in the disc
\citep{Devereux+92}. 
HST UV images show that the present star formation in the ring of NGC\,3351 is
arranged in complexes of diameters between 1.4 and 2.0~arcsec, 
made up of several high-surface brightness knots a few parsec in size
embedded in a more diffuse component \citep{Colina+97}.
A circumnuclear SFR of 0.38\,M$_\odot$\,yr$^{-1}$ was inferred by
\cite{Elmegreen+97} from near infrared (IR) photometry in  the J and K bands,
and a value of 0.24\,M$_\odot$\,yr$^{-1}$ was derived by
\cite{Planesas+97} from the H$\alpha$ emission.
From CO observations, these authors estimated a mass of molecular gas
of 3.5\,$\times$\,10$^{8}$\,M$_\odot$ inside a circle of 1.4~kpc in diameter.
According to \cite{Jogee+05}, NGC\,3351 hosts 5.3\,$\times$\,10$^8$\Msun of
molecular hydrogen, with most of the emission coming from the inner
$\sim600$\,pc. 
Within this radius, a SFR of 0.5\,M$_\odot$\,yr$^{-1}$ can be estimated from
the Br$\gamma$ luminosity given by \cite{Puxley+90}, whereas the non-thermal
component of the radio continuum emission favours a lower value of
0.3\,M$_\odot$\,yr$^{-1}$ \citep{Jogee+05}. 
\cite{vandeVen+09} estimated that the molecular mass inside the circumnuclear
ring is about 
5\,$\times$\,10$^7$\Msun, and \cite{Leroy+09} found a well-defined molecular ring
with a diameter of about 50\,arcsec, located at the end of the stellar bar
\citep[with a semimajor axis of 47\,arcsec; see e.g.][]{Martin95}. These CO
structures are compatible with the excesses observed by \cite{Regan+01} and
\cite{Helfer+03}. A small CO bar perpendicular to the large-scale stellar bar
was also observed by \cite{Devereux+92} and \cite{Helfer+03}. 

NGC\,3351 is part of the SINGS Legacy Science Program sample
\citep{Kennicutt+03}, and it has been included in a wide variety of
statistical studies. The SINGS data set provides broadband imaging, as well as
low- and high-resolution spectral maps in the IR. The integrated spectral
energy distribution (SED) of NGC\,3351 from 0.15 to 850~$\mu$m has been
derived by \cite{Dale+07} by combining the {\it Spitzer} data with those from
2MASS and {\it IRAS} in the IR, GALEX in the UV, and other data in the
optical, modelling the dust and stellar contributions. 
They concluded that
most of the sample's spectral variations stem from two underlying components,
one representative of a galaxy with a low IR-to-UV ratio and one
representative of a galaxy with a high IR-to-UV ratio, as given
by a principal component analysis of the sample.
\cite{Draine+07} used the IRAC and MIPS observations to estimate the total
dust mass (10$^{7.46}$\,\Msun) and the fraction of the dust mass contributed
by PAHs (3.2 per cent). They found that the dust properties
of spiral galaxies resemble those in the local environment of the Milky Way, with
similar dust-to-gas ratio and similar PAH abundance. \cite{Munoz+09} found
that NGC\,3351 shows significantly larger IR concentration indices than
the remaining galaxies of the same morphological type of the SINGS
sample.

We have studied the circumnuclear region of NGC\,3351 in two previous works.
In \cite{Diaz+07}, the physical conditions of the gas were determined
for 7 CNSFRs. 
We developed a semi-empirical method for the derivation of chemical abundances
in the high-metallicity regime, obtaining values consistent with solar within
the errors and comparable to those found in high-metallicity disc \HII regions
using direct measurements of electron temperatures. An analysis of the
kinematics of the gas and stars in 5 CNSFRs and the nucleus of NGC\,3351 is
presented in \cite{Hagele+07}. We derived the dynamical masses (between
$4.9\times10^6$ and $4.3\times10^7$\,M$_\odot$) for the stellar clusters in
these regions, and found that the CNSFR complexes, with sizes of about 100 to
150\,pc in diameter, are seen to be composed of several individual star
clusters (with sizes between 1.7 and 4.9\,pc on a HST image).
The radial velocity curve shows deviation from circular motions for the
ionised hydrogen, consistent with infall towards the central regions of
the galaxy at a velocity of about 25\,km\,s$^{-1}$ \citep[in agreement
  with][]{Rubin+75,Dicaire+08}. 
These last authors also point out that the H$\alpha$ velocity field is fairly
regular outside the  central bar, and \cite{Jogee+05} found that the velocity
field of the molecular material in the inner 500\,pc is generally dominated by
circular motions, with some evidence of weaker non-circular streaming. 

In this paper, we will look for radio-continuum signatures of massive star
formation in NGC\,3351, both in the nucleus (in particular, the CNSFRs) and
the galaxy disk. 
Section~\ref{secObs} describes the observations and the data reduction.
The spatial distribution of radio sources and their observed properties are
discussed in Section~\ref{secResults}, and Section~\ref{secDiscus} is devoted
to the interpretation of these results in terms of star formation activity. 
Our main conclusions are briefly summarised in Section~\ref{secConclus}.


\begin{table*}
\caption{Sources around NGC\,3351 detected by MERLIN. The first column
gives the adopted reference name for each source (in brackets the
adopted names for the three closest sources to the centre of
NGC\,3351), the second column gives the FIRST counterpart name (within
1\,arcsec of the MERLIN source in all cases) and the third column
gives the radial offset of the MERLIN sources from the pointing
centre. The remaining columns give the MERLIN source properties: positions,
peak flux densities, integrated flux densities, and their corresponding
errors. These measurements for R1, R2 and R3 were obtained by fitting a single
component; see Table \ref{ext-sources}\ for further details of these
sources.
}
\label{tab:brights}
\begin{tabular}{lcrccrrrrr}
\hline
Source & FIRST & Radius& R.A. & Dec. & $\sigma_{\rm {RA}}$ & $\sigma_{\rm {Dec}}$ & Peak & Int. &$\sigma_{\rm {rms}}$\\
 & name & (arcmin) & \multicolumn{2}{c}{(J2000)}& (mas)& (mas)&(mJy b$^{-1}$)&(mJy)&(mJy b$^{-1}$)\\
\hline
 1    &  J104252.6+114853 &17.283 & 10 42 52.5564 & +11 48 54.157 & 47 & 76 &  0.31 &   1.78 & 0.08 \\
 2    &  J104253.2+114848 &17.076 & 10 42 53.3084 & +11 48 48.482 & 27 & 44 &  0.55 &   1.91 & 0.08 \\
 3    &  J104254.6+114841 &16.727 & 10 42 54.6482 & +11 48 41.204 & 24 & 38 &  0.63 &   5.33 & 0.08 \\
 4    &  J104301.1+115230 &17.228 & 10 43 01.1845 & +11 52 30.364 & 27 & 44 &  0.54 &   1.28 & 0.08 \\
 5    &                   &15.086 & 10 43 12.8952 & +11 52 35.685 & 48 & 78 &  0.31 &   0.73 & 0.08 \\
 6    &  J104320.5+114301 &9.1344 & 10 43 20.5287 & +11 43 01.933 & 10 & 16 &  1.48 &   5.00 & 0.08 \\
 7    &  J104328.4+113723 &8.6480 & 10 43 28.4280 & +11 37 23.613 &  5 & 10 &  2.43 &   5.53 & 0.08 \\
 8    &                   &5.2228 & 10 43 36.7801 & +11 43 15.556 & 21 & 34 &  0.72 &   1.65 & 0.08 \\
 9    &  J104342.7+115209 &10.575 & 10 43 42.7584 & +11 52 09.373 &  2 &  4 &  4.33 &  11.82 & 0.08 \\
10    &                   &2.8629 & 10 43 47.3236 & +11 40 54.775 & 35 & 57 &  0.42 &   0.81 & 0.08 \\
11    &  J104349.1+113628 &6.1266 & 10 43 49.1642 & +11 36 28.452 &  3 &  5 &  2.74 &   4.00 & 0.09 \\
12    &                   &1.9497 & 10 43 49.8033 & +11 41 58.791 & 43 & 70 &  0.43 &   0.85 & 0.08 \\
13    &                   &2.8251 & 10 43 50.6611 & +11 39 59.677 & 35 & 56 &  0.42 &   0.78 & 0.08 \\
14(R1)&  J104355.2+114129 &0.9541 & 10 43 55.2618 & +11 41 29.337 &  2 &  4 &  4.11 &   4.08 & 0.07 \\
15(R2)&  J104359.5+114332 &1.3864 & 10 43 59.5464 & +11 43 32.638 &  3 &  5 &  3.53 &  22.04 & 0.07 \\
16(R3)&  J104401.4+114323 &1.4793 & 10 44 01.4482 & +11 43 23.623 & 21 & 18 &  1.38 &  15.93 & 0.06 \\
17    &                   &2.9359 & 10 44 08.4135 & +11 40 54.837 & 38 & 58 &  0.39 &   1.09 & 0.08 \\
18    &                   &3.0888 & 10 44 10.3120 & +11 42 08.643 & 35 & 57 &  0.42 &   0.50 & 0.08 \\
19    &  J104417.4+114614 &6.2824 & 10 44 17.4533 & +11 46 14.678 &  1 &  1 & 62.77 & 125.40 & 0.21 \\
20    &                   &6.6425 & 10 44 17.5254 & +11 46 46.157 & 15 & 24 &  0.96 &   1.79 & 0.09 \\
21    &  J104423.4+114458 &6.8677 & 10 44 23.4286 & +11 44 58.340 &  3 &  3 & 14.05 &  32.41 & 0.32 \\
22    &  J104429.2+113811 &8.7224 & 10 44 29.2655 & +11 38 11.349 &  7 &  6 &  4.05 &  13.05 & 0.14 \\
23    &  J104430.6+113811 &9.0252 & 10 44 30.6493 & +11 38 11.184 &  1 &  1 & 21.12 &  88.18 & 0.13 \\
\hline
\end{tabular}
\end{table*}



\begin{figure*}
\centering
  \includegraphics[width=1\textwidth]{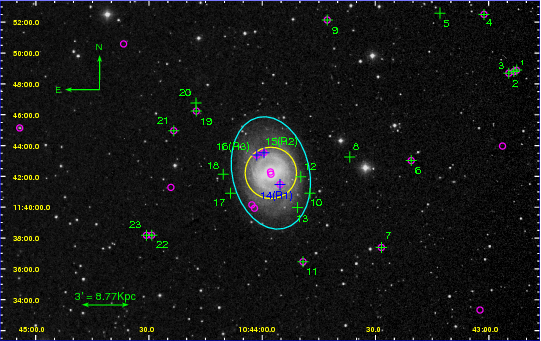}
  \caption[]{Distribution of the MERLIN sources in a
  35\,$\times$\,22\,arcmin$^2$ field centred at $\alpha_{\rm 
  2000}$\,=\,10$^{\rm h}$\,43$^{\rm m}$\,57\fs7, $\delta_{\rm
  2000}$\,=\,+11\degr 42\arcmin 14\farcs0 superimposed in an H$\alpha$ image
  from Palomar 48-inch Schmidt telescope. Crosses and 
  small circles indicate the MERLIN and FIRST sources, respectively. The
  MERLIN sources are labelled according to Table~\ref{tab:brights}. The big 
  circle indicates the central 100\,arcsec (in radius) of NGC\,3351, and the
  ellipse is the D$_{25}$ isophote of this galaxy as defined by
  \citet{deVaucouleurs+91}. The orientation is north up, east to the
  left. [{\it See the electronic edition of the Journal for a   
  colour version of this figure.}]}
\label{all-source-dist}
\end{figure*}


\section{Observations and data reduction}
\label{secObs}

We observed NGC\,3351 at 1.42\,GHz on 2008 Jan 26-28 using 6 antennas of
the Multi-Element Radio Linked Interferometer Network (MERLIN). The Cambridge
telescope has a diameter of 32\,m and provides the longest baselines of up
to 217\,km; the other five antennas have effective diameters of 25
meters. The field centre was at right ascension $10^{\rm h}\,43^{\rm
  m}\,57\fs7000$, declination +11\degr 42\arcmin 14\farcs000 (J2000). We used
the 
point-like QSO B1023+131, at a separation of less than $5$\degr, as the phase
reference 
source, with a duty cycle of about 6.5:1.5 min on the target and phase
reference, respectively.
The position used for B1023+131 ($10^{\rm h}\,25^{\rm m}\,56\fs2854$, +12\degr
53\arcmin 49\farcs024) was taken from 
\cite{Browne98}. We used a 4\,s integration time.  After flagging bad data,
the total, usable time on NGC\,3351 was 15.4 hours, which is expected to give a
noise level of  $\sigma_{\rm {rms}} = 53~\mu$Jy~beam$^{-1}$ under good
conditions.
The point-like QSO OQ208 was used as a bandpass calibration source and to set
the flux scale on all baselines.
Its flux density was found to be $0.955\pm0.008$\,Jy by comparison with 3C286
on the shorter baselines, where the latter is unresolved and has a flux
density of 14.73\,Jy at 1.42\,GHz \citep{Baars77}.

We followed the standard MERLIN continuum data reduction procedure (MERLIN
User Guide, Diamond et
al. 2003\footnote{http://www.merlin.ac.uk/user\_guide/OnlineMUG/}), using the
{\sc aips} package\footnote{http://www.aips.nrao.edu/} for calibration and
imaging. We applied the flux scale, bandpass and time-variable phase and
amplitude solutions from the reference sources to NGC\,3351 and split out the
data. We retained 15 1-MHz channels. This spectral resolution gives the most
stringent restriction on the field of view, of about 10 percent smearing at
140\,arcsec radius. We used separate facets of a few arcmin in all mappings
in order to avoid distortions due to sky curvature and natural weighting,
giving a beam size of $0.29 \times 0.17$~arcsec with a position angle of
23\degr. We searched for confusing 
sources by making a grid of 1681 100-arcsec maps, without cleaning, over a
1-degree square field.  We then re-imaged and cleaned all fields containing
sources brighter than 2\,mJy, using the clean components as a model for
self-calibration of the whole dataset.  We carried out further rounds of
imaging and self-calibration, testing different strategies such as
establishing that restricting self-calibration to the inner sources gave
poorer results. We established that sidelobes fell to $\la5$ percent of the
peak at distances $\ga0.5$ arcmin from a source and identified all sources
brighter than 1 mJy within the central 12 arcmin. Fainter sources should not
produce sidelobes above the expected $\sigma_{\rm rms}$ in the central region.

\section{Results}
\label{secResults}

\begin{figure}
\centering
\includegraphics[width=0.47\textwidth]{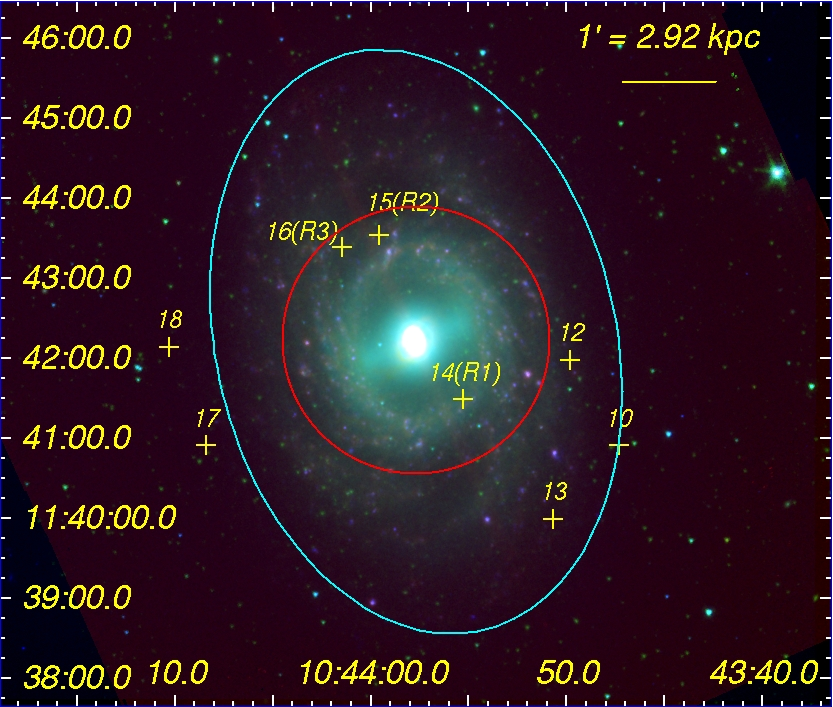}
\caption{Superposition of the Spitzer images at 8.0 $\mu$m (red) and 3.6
  $\mu$m (green) with $H_{\alpha}$ (blue) emission distribution from
  NGC\,3351. The yellow crosses indicate the radio sources projected over the
  galaxy, 
  labelled according to Table~\ref{tab:brights}. The big red circle indicates
  the central 100\,arcsec (in radius) of NGC\,3351, and the cyan ellipse is 
  the D$_{25}$ isophote of this galaxy as defined by
  \citet{deVaucouleurs+91}. The orientation is north up, east
  to the left.}
 \label{rgb}
\end{figure}

Table~\ref{tab:brights} gives the properties of the 23 radio-continuum sources
at 1.42~GHz found by MERLIN within a $35\times22$~arcmin region around
NGC\,3351. The first column identifies the sources. The three objects that are
closest to the centre (14, 15, and 16) will be referred to as R1, R2, and R3
in what follows. 
Fifteen of the MERLIN sources have a counterpart in the VLA FIRST (Very Large
Array, Faint Images of the Radio Sky at Twenty-cm) survey catalogue
\citep{White+97} within 1~arcsec radius; names of the FIRST counterparts are
quoted in column~2. 
Column~3 shows the radial offset from the MERLIN pointing centre, and
columns~4 to 7 give the coordinates of each source and their corresponding
errors. Peak surface brightness, integrated flux densities, and local noise
levels $\sigma_{\rm{rms}}$ are listed in columns 8, 9, and 10, respectively. 
The distribution of the MERLIN sources is shown in Fig.~\ref{all-source-dist}
overplotted on an H$\alpha$ image acquired with the Palomar 48-inch Schmidt
telescope and obtained from the Digitized Sky
Survey\footnote{http://stdatu.stsci.edu/cgi-bin/dss\_form} (DSS) through the
NASA/IPAC Extragalactic Database\footnote{http://nedwww.ipac.caltech.edu/}
(NED). 

We found six sources inside the D$_{25}$ isophote whose
intensity peaks are brighter than the local 6 $\sigma_{\mathrm {rms}}$
confidence limit of $310-340~\mu$Jy\,beam$^{-1}$, corresponding to a
brightness temperature of $\sim$\,1000\,K. Outside this ellipse (D$_{25}$),
the noise increases, reaching about 0.07--0.08\,mJy\,beam$^{-1}$ in the
regions out to $\sim$15-20 arcmin radius which are affected by bandwidth and
other smearing, making accurate deconvolution difficult. We visually examined
each candidate source, rejecting those which seemed spurious. Artefacts did
not exceed 6\,$\sigma_{\rm rms}$ within the primary beam FWHM, except for
clearly-recognisable sidelobes reaching 10\,$\sigma_{\rm {rms}}$ within $\sim
1$~arcmin of some of the brightest outlying sources In the outer part of the
field, we retained fifteen sources exceeding 10 $\sigma_{\mathrm {rms}}$
($\sim700-800$ mJy beam$^{-1}$) and two fainter sources (5 and 18) which are
conspicuously isolated from any contaminating sidelobes. 
The distribution of radio sources close to the centre of NGC\,3351 is shown
with crosses in Figure~\ref{rgb}. 
This false colour image is made using two Spitzer images at 8.0 and
3.6\,$\mu$m (red and green channels, respectively) and a continuum-subtracted
H$\alpha$ image (blue channel) from CTIO 4.0\,m Mosaic2 (correction to the
celestial coordinates of this image was provided by Douglas Swartz, private
communication).

MERLIN detected 15 (about 63\,\%) of the 24 sources listed in the
FIRST survey catalogue within the field of view shown in
Fig.~\ref{all-source-dist}. The area of the MERLIN beam is $\le1\%$ of
the FIRST resolution of $\sim 5$~arcsec, giving a brightness
sensitivity for MERLIN $>1000$ K compared with 20-30 K for FIRST.  In
addition, the shortest MERLIN baselines correspond to a maximum
imageable angular scale of $\approx2$ arcsec. All the non-detected
sources are fainter than 7 mJy per FIRST beam.
 On the other hand, we have discovered 8 new sources that were not
previously catalogued in FIRST, due to our greater sensitivity to
compact emission. The position errors quoted in
Table~\ref{tab:brights} are the noise-based relative errors of the
source peaks. Astrometric uncertainties also arise from the assumed
positions of the phase reference source ($\sim$\,3\,mas) and the
MERLIN antennas ($\sim$\,10\,mas), as well as from the phase
reference-target separation ($\sim$\,5\,mas), giving a total
astrometric uncertainty of about 12\,mas. The sources at increasing
distances from the centre, over a few arcmin, may suffer from
additional errors in the peak position due to smearing.

The best images of the central region were achieved by using all sources
identified in a 1-degree region for self-calibration and subtracting the
clean components for sources further than 2~arcmin from the centre.
After all calibration and subtraction, we made a cleaned image of the region within
100\,arcsec of the centre. Using natural weighting gave $\sigma_{\rm
{rms}} = 57\,\mu$Jy\,beam$^{-1}$. We tried giving the longer baselines lower
weight (tapering) and used a restoring beam of $0.5 \times 0.5$~arcsec,
giving $\sigma_{\rm {rms}} = 52~\mu$Jy.
Although these levels are close to the predicted noise (allowing for the low
declination of NGC\,3351), even the best images contained ripples of magnitude
$\sim 6\,\sigma_{\rm {rms}}$ due to the severe distortions introduced by
bright sources at large offsets (which have peaks of up to a few Jy in FIRST)
that could not be adequately subtracted or cleaned out. 


\begin{figure*}
\centering
\includegraphics[width=.33\textwidth,angle=0]{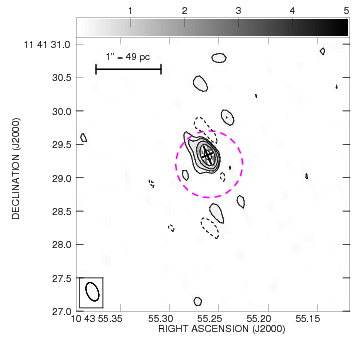}
\includegraphics[width=.33\textwidth,angle=0]{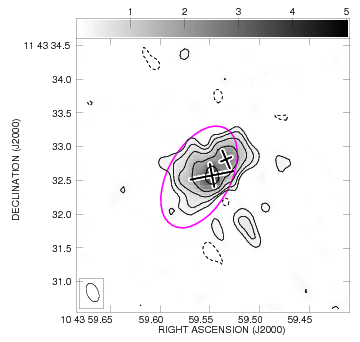}
\includegraphics[width=.33\textwidth,angle=0]{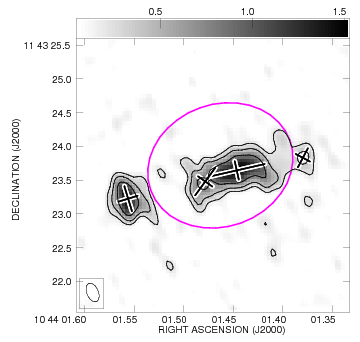}
\caption{Radio continuum MERLIN map of the sources R1, R2 and
  R3 (left, middle and right panel, respectively). The scale is the same in
  all the panels. Level contours correspond to
  -1, 1, 2, 4...\,$\times$\,0.2\,mJy\,beam$^{-1}$. The crosses represent
  the major and minor axes of the deconvolved components fitted to the MERLIN
  images and the ellipses correspond to the FIRST deconvolved source
  shapes (the dashed circle represents the unresolved source). The orientation
  is north up, east to the left. [{\it See the electronic edition of the
  Journal for a colour version of this figure.}]} 
  \label{extended-sources}
\end{figure*}



\begin{figure*}
\centering
\includegraphics[width=.4\textwidth,angle=0]{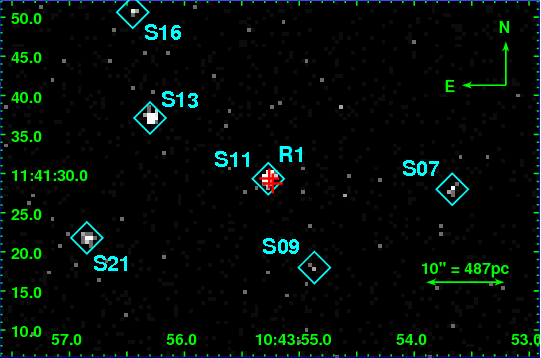}
\hspace{0.1cm}
\includegraphics[width=.4\textwidth,angle=0]{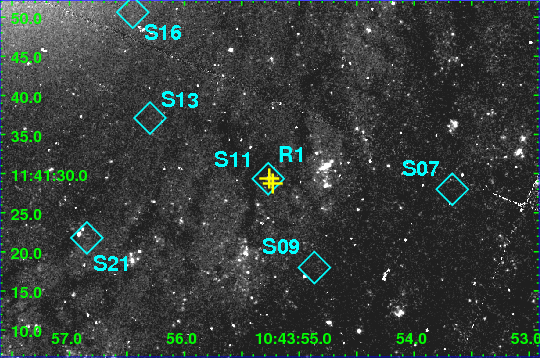}\\
\vspace{0.3cm}
\includegraphics[width=.4\textwidth,angle=0]{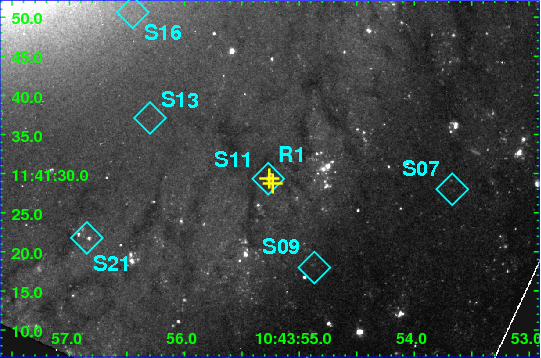}
\hspace{0.1cm}
\includegraphics[width=.4\textwidth,angle=0]{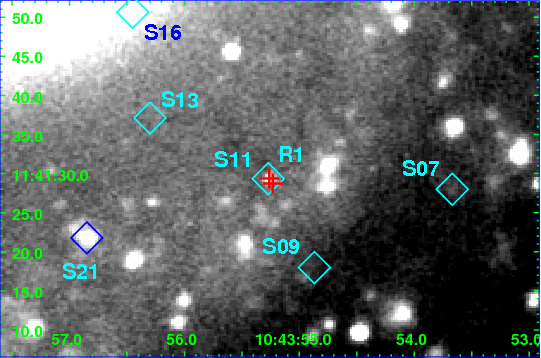}\\
\vspace{0.3cm}
\includegraphics[width=.4\textwidth,angle=0]{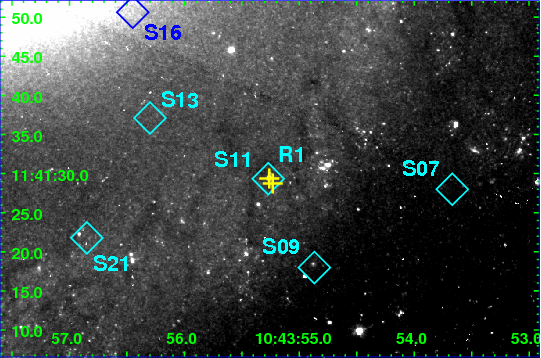}
\hspace{0.1cm}
\includegraphics[width=.4\textwidth,angle=0]{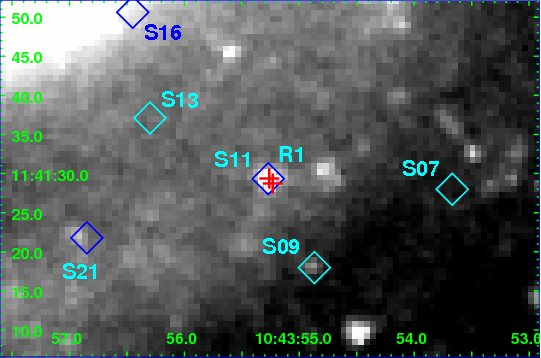}\\
\vspace{0.3cm}
\includegraphics[width=.4\textwidth,angle=0]{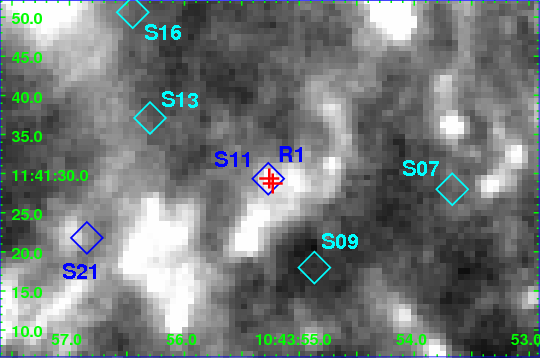}
\hspace{0.1cm}
\includegraphics[width=.4\textwidth,angle=0]{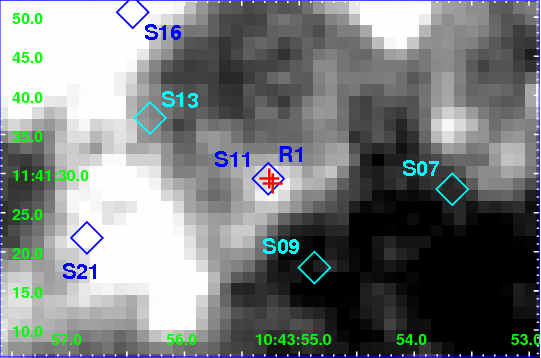}\\

\caption{Images of 1.15\arcmin\,$\times$\,0.75\arcmin\ around R1. From top,
  left to right: Chandra 0.1-8\,keV, HST-WFPC2 F439W 
  (wide B), HST-WFPC2 F606W (wide V), CTIO H$\alpha$, HST-WFPC2 F814W (wide
  I), Spitzer-IRAC 4.5\,$\mu$m, Spitzer-IRAC 8\,$\mu$m, Spitzer-MIPS
  24\,$\mu$m. In all panels crosses show the two detected knots of R1,
  diamonds represent the x-rays sources detected by Chandra with the
  nomenclature defined by \citet{Swartz+06}. In all the cases the orientation
  is north up, east to the left. [{\it See the electronic edition of the
  Journal for a colour version of this figure.}]}
\label{R1-multiw}
\end{figure*}



\begin{figure*}
\centering
\includegraphics[width=.4\textwidth,angle=0]{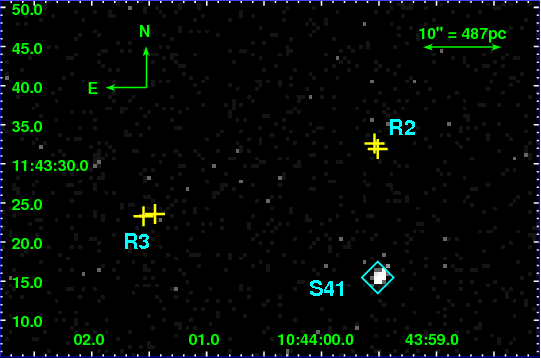}
\hspace{0.1cm}
\includegraphics[width=.4\textwidth,angle=0]{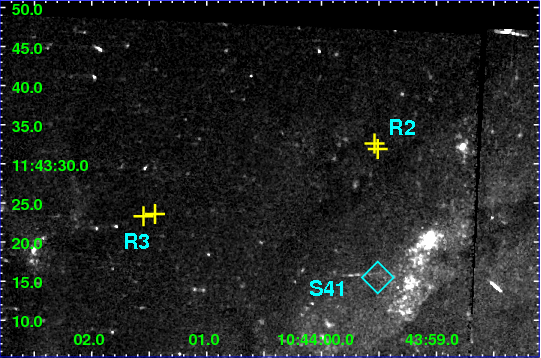}\\
\vspace{0.3cm}
\includegraphics[width=.4\textwidth,angle=0]{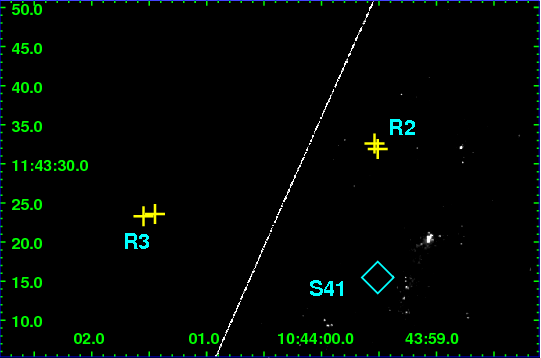}
\hspace{0.1cm}
\includegraphics[width=.4\textwidth,angle=0]{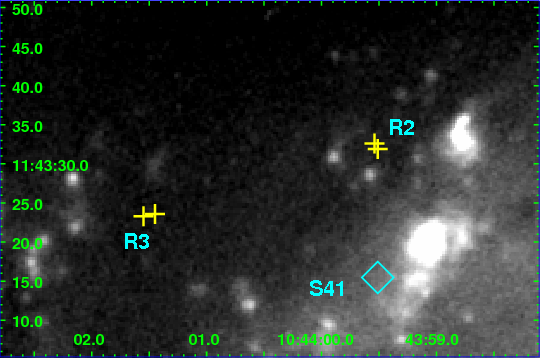}\\
\vspace{0.3cm}
\includegraphics[width=.4\textwidth,angle=0]{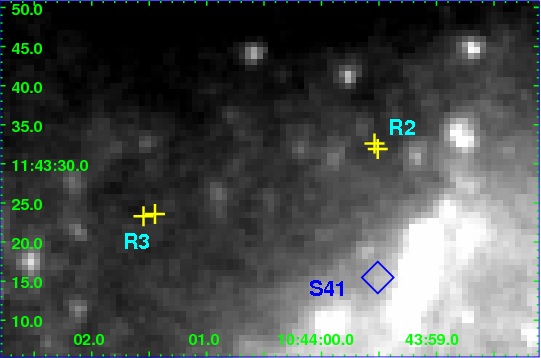}
\hspace{0.1cm}
\includegraphics[width=.4\textwidth,angle=0]{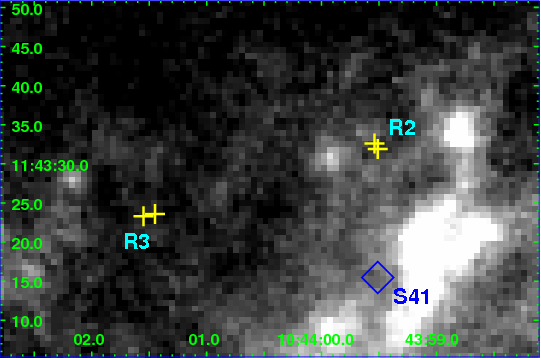}\\
\vspace{0.3cm}
\includegraphics[width=.4\textwidth,angle=0]{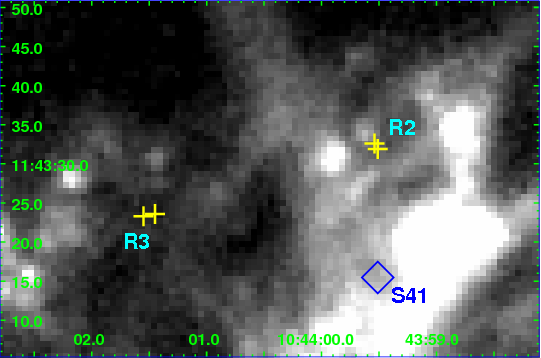}
\hspace{0.1cm}
\includegraphics[width=.4\textwidth,angle=0]{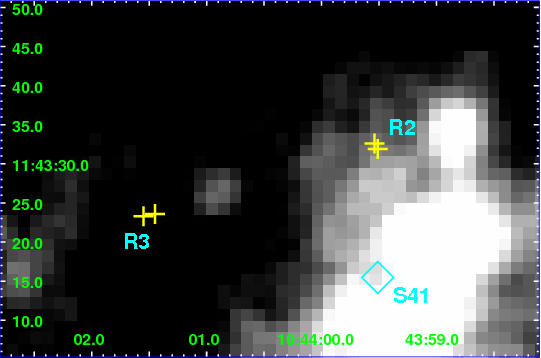}\\

\caption{Images of 1.15\arcmin\,$\times$\,0.75\arcmin\ around R2 and R3. From
  top, left to right: Chandra 0.1-8\,keV, HST-WFPC2 F450W 
  (wide B), HST-WFPC2 F606W (wide V), CTIO H$\alpha$, Spitzer-IRAC 3.6\,$\mu$m,
  Spitzer-IRAC 5.8\,$\mu$m, Spitzer-IRAC 8\,$\mu$m, Spitzer-MIPS 
  24\,$\mu$m. In all panels crosses show the locations of R2 and R3,
  diamonds represent the x-rays sources detected by Chandra with the
  nomenclature defined by \citet{Swartz+06}. In all the cases the orientation
  is north up, east to the left. [{\it See the electronic edition of the
  Journal for a colour version of this figure.}]}
\label{R2+R3-multiw}
\end{figure*}


\begin{table*}
\begin{center}
\caption{Radio continuum parameters at 1.42\,GHz from the extended
sources. The first column gives the adopted reference name. From column 2 to 6 
we list the peak position, the peak intensity, integrated flux density and
sizes (major axis $\times$ minor axis, position angle) of the different knots of
these sources. The last two columns give the values listed in the FIRST survey
\citep{White+97}.} 
\begin{tabular}{lccccc|cc}
\hline
Region&\multicolumn{5}{c|}{MERLIN}&\multicolumn{2}{c}{FIRST}\\
     &    R.A.  & Dec.           & Peak   &   Int.         & Size                &Int    & Size                  \\
&\multicolumn{2}{c}{(J2000)}&(mJy beam$^{-1}$)&(mJy)        &(arcsec$\times$arcsec,deg.)&(mJy)&(arcsec$\times$arcsec,deg.)\\
\hline
R1   &10 43 55.2618 & +11 41 29.337 &   4.11 & $ 4.08\pm0.06$ & (compact)           & 1.91  &       (compact)       \\
R2   &10 43 59.5480 & +11 43 32.572 &   3.82 & $23.76\pm0.37$ &$0.65\times0.35$,101 & 22.12 & $1.64\times0.93$,151.3\\	
     &10 43 59.5338 & +11 43 32.810 &   3.53 & $ 3.53\pm0.06$ &(unresolved)         &  --   &      --               \\
R3   &10 44 01.4465 & +11 43 23.646 &   1.50 & $10.57\pm0.41$ &$0.85\times0.28$,103 & 23.19 & $2.19\times1.81$,109.6     \\
     &10 44 01.5560 & +11 43 23.228 &   1.46 & $ 4.93\pm0.23$ &$0.39\times0.30$,17  &  --   &      --               \\
     &10 44 01.4783 & +11 43 23.482 &   0.92 & $ 3.84\pm0.27$ &$0.51\times0.26$,144 &  --   &      --               \\
     &10 44 01.3798 & +11 43 23.832 &   0.47 & $ 1.29\pm0.19$ &$0.32\times0.24$,152 &  --   &      --               \\
\hline
\end{tabular}
\label{ext-sources}
\end{center}
\end{table*}

Three sources, abbreviated by us as R1, R2 and R3, were found within the
central 100~arcsec in radius (see Figures~\ref{rgb}
and~\ref{extended-sources}). All of them are brighter than 1\,mJy, and both R2
and R3 are significantly extended. R1 seems to be very compact, and MERLIN
measured a higher flux density (4.1\,mJy) than FIRST (1.9\,mJy), suggesting
not only that this source is smaller than the MERLIN beam, but also that it is
compact enough to be highly variable, having approximately doubled in flux
between the date of the FIRST survey and that of the MERLIN 
observations. The FITS header of the
observations in the VLA archive shows that they were made in Dec 1999
and Jan 2000. That means that the size of R1 is constrained to be about 2.4
pc, corresponding to a light-travel diameter of 8 yr.

As can be appreciated in Figure~\ref{extended-sources}, these sources display
several faint, yet observable, substructures. R1 displays a relatively compact
main knot, and the other features are too faint to be sure they are real. R2
has an unresolved knot to the southwest, not necessarily compact, but its
components are too faint and irregular, or too blended, to measure their
extent reliably. R3 shows three fainter knots. Although the intensity peak (in
mJy per beam) of the brightest one is almost equal to that of the main knot,
its integrated flux is less than a half. Table~\ref{ext-sources} lists
the coordinates, peak surface brightness, integrated flux densities, and
sizes for each knot, compared to the analogous values given in the FIRST
survey \citep{White+97}. 
There are a few other features extended along the beam angle, but they seem to
be artefacts and thus we have not listed them in the table. All fluxes and
sizes, as well as their corresponding errors, have been estimated from the 4,
6 and 8\,$\sigma_{\rm {rms}}$ level contours. Sizes are given after
deconvolution of the restoring beam, and their errors are smaller than
0.02~arcsec (5\,\degr\ in position angle) for sources brighter than
1\,mJy. For fainter sources, the errors increase to 0.05\,arcsec and
20\,\degr\ for sizes and position angles, respectively. {The discrepancy
between the two quoted position angles of R2 between the FIRST and MERLIN
measurements (see middle panel of Figure~\ref{extended-sources}) is partly due
to the fact that the MERLIN PA is for the brighter peak only, whilst the FIRST
PA is derived from a fit to the entire patch of emission. The position angle
between the two MERLIN peaks is 136 deg, which agrees with the FIRST PA within
the errors. FIRST does not give individual position angle errors. We have 
downloaded the image from the FIRST archive\footnote{http://sundog.stsci.edu/}
and fitting R2 and R3 we have 
obtained an uncertainty of 14 and 18 deg, respectively.}

\begin{table*}
\caption
{Main properties of each region: diameter, \Ha luminosity, electron density and
temperature, filling factor, thermal optical depth and flux density, star
formation rate, total number of SNRs, expected number above our detection
threshold, and total synchrotron flux density.
}
\label{tab:regions}
\begin{tabular}{@{}l@{\hspace{0.3cm}}c@{\hspace{0.3cm}}c@{\hspace{0.3cm}}c@{\hspace{0.3cm}}c@{\hspace{0.3cm}}c@{\hspace{0.3cm}}c@{\hspace{0.3cm}}c@{\hspace{0.3cm}}c@{\hspace{0.3cm}}r@{\hspace{0.3cm}}c@{\hspace{0.3cm}}r@{}}
\hline
Region
& $D$ & $L(\Ha)$ & $n_{\rm e}$ & $T_{\rm e}$ & $f$
& $\tau_{\rm 1.42\,GHz}^{\rm brems}$
& $S_{\rm 1.42\,GHz}^{\rm brems}$
& SFR & $N_{\rm SNR}$
& $N_{\rm SNR}^{( 0.3~{\rm mJy} )}$
& $S_{\rm 1.42\,GHz}^{\rm synchr}$
\\
& (arcsec) & ($10^{39}$\,erg\,s$^{-1}$) & (cm$^{-3}$) & ($10^4$ K) & &
& (mJy) & ($M_\odot$~yr$^{-1}$) & & & (mJy)~~
\\\hline
CNSFRs  & 1.00 & 5.43 & 360 & 0.48 & 0.021 & 0.212 & 0.30 & 0.060 & 4.4 & 0.119 & 8.38 \\
        & 1.00 & 2.83 & 440 & 0.45 & 0.007 & 0.165 & 0.16 & 0.031 & 2.3 & 0.062 & 4.37 \\
        & 1.00 & 7.00 & 430 & 0.45 & 0.018 & 0.296 & 0.36 & 0.077 & 5.7 & 0.154 & 10.80 \\
        & 0.80 & 2.32 & 310 & 0.44 & 0.022 & 0.145 & 0.13 & 0.026 & 1.9 & 0.051 & 3.58 \\
        & 0.60 & 1.03 & 360 & 0.54 & 0.021 & 0.106 & 0.06 & 0.011 & 0.8 & 0.023 & 1.59 \\
        & 1.50 & 0.75 & 360 & 0.55 & 0.001 & 0.034 & 0.05 & 0.008 & 0.6 & 0.017 & 1.16 \\
        & 1.00 & 1.80 & 410 & 0.46 & 0.005 & 0.114 & 0.10 & 0.020 & 1.5 & 0.040 & 2.78 \\
        & 0.60 & 0.54 & 400 & 0.50 & 0.008 & 0.079 & 0.03 & 0.006 & 0.4 & 0.012 & 0.83 \\
Centre  & 24.00 & 23.60 & 10 & 1.00 & 0.016 & 0.001 & 2.14 & 0.260 & 19.2 & 0.519 & 36.41 \\
Disk    & 400.00 & 45.50 & 1 & 1.00 & 0.001 & 0.000 & 4.13 & 0.500 & 125.1 & 1.001 & 70.20 \\
\hline
\end{tabular}
\end{table*}

R1 is the only source that has a clear counterpart at other wavelengths, from X-rays to radio.
It has already been observed and identified as an X-ray point source using Chandra data \citep[S11 in the catalogue of][]{Swartz+06}.
Galex at near- and far-UV, as well as Spitzer-MIPS at 70 and 160\,$\mu$m in the far-IR, do not show a clear counterpart due to their lower spatial resolution, leading to confusion and blending with nearby
objects.
Images of the region around R1 in different bands are shown in
Figure~\ref{R1-multiw}: X-rays from the Chandra Data
Archive\footnote{http://cda.harvard.edu/chaser/mainEntry.do}, H$\alpha$ and IR
images from NED, HST-WFPC2 images from the Hubble Legacy
Archive\footnote{http://hla.stsci.edu/hlaview.html} (HLA) and the Multimission
Archive at STScI\footnote{http://archive.stsci.edu/} (MAST). 
The X-ray image has been built by combining the three $\sim 40$~ks images
available in the archive using the {\sc ciao} 3.4 software
\citep{Fruscione+06}. For R2 and R3, we find no obvious counterpart at any
wavelength (see Figure~\ref{R2+R3-multiw}). In fact, they seem to be located
in regions devoid of emission in other bands. There is no correlation between
these MERLIN radio sources and other catalogues, such as the atlas of \HII
regions by \citet{Hodge+83} or \citet{Bradley+06}. A general search in NED for
entries around the positions of R2 and R3 also yielded negative results. They
do not have satisfactory cross-identifications in SPECFIND \citep{Vollmer+10},
although either or both could be associated with 611 and 4850\,MHz sources
observed at lower resolution. The high flux densities observed at 611\,MHz
hint at these sources having a non-thermal origin. 

We found no further sources brighter than $\sim$\,6\,$\sigma_{\rm {rms}}$ in
the central region. In particular, the two nuclear sources catalogued in FIRST
are resolved out in the present data. At larger radii, MERLIN observed 3
additional sources above the 6\,$\sigma_{\rm {rms}}$ limit that are projected
inside the D$_{25}$ isophote of NGC\,3351. These sources (entries 10, 12, and
13 in Table~\ref{tab:brights}, located at 2--3~arcmin from the centre of the
observed field) were not detected in the FIRST survey nor in the VLA
observations reported in \citet{Paladino06}.

\section{Discussion}
\label{secDiscus}

Early-type spiral galaxies such as NGC\,3351 provide an ideal laboratory to
study the very different environments in which star formation may take place,
from the peculiar conditions of the CNSFRs
\citep[e.g.][and references therein]{Hagele+09,Hagele+10} to the outskirts of
the galaxy disk. 
Maps of radio-continuum emission can reveal the presence of SNRs and \HII
regions through their synchrotron and thermal emission, respectively. 
These are closely related to the evolution of massive stars and are,
therefore, excellent tracers of the recent star formation activity within the
galaxy \citep[see e.g.][]{Muxlow+94}. 
Moreover, young massive star-forming regions have very characteristic
signatures in the radio band, providing strong constraints on the age of the
most recent starbursts \citep[see e.g.][]{Rosa-Gonzalez+07a}.

We have divided the galaxy in several regions, corresponding to different star
formation regimes: the centre, which we define as the central 600~pc
(12~arcsec radius) features the highest densities and metallicities, $Z \sim
Z_\odot$. 
The CNSFR complexes identified in \citet{Planesas+97} have been considered
separately from the diffuse nuclear star formation. 
Finally, the galaxy disk ($r>600$ pc, $r>12$~arcsec) represents a more
typical, low-density environment with low to moderate metallicities.
The main properties (diameter, \Ha luminosity, electron density and
temperature, filling 
factor, thermal optical depth and flux density, star formation rate,
total number of SNRs, expected number above our detection threshold, and total
synchrotron flux density) of each region are summarised in
Table~\ref{tab:regions}.
The eight CNSFRs are listed in the same order as in \citet{Planesas+97}.

Diameters ($D$) of the CNSFR complexes are estimated from the HST F606W image,
and \Ha luminosities [$L(\Ha)$] are taken from \citet{Planesas+97},
corrected for internal extinction using the
colour excess [E(B-V)] estimated by \citet{tesisdiego} and assuming the
Galactic extinction law of \citet{Miller+72} with $R_v$\,=\,3.2
\citep{HagelePhD}.
For the rest of the nuclear region (excluding the CNSFRs), the star formation
rate derived from the Br$\gamma$ luminosity \citep{Puxley+90} has been
converted to an \Ha 
luminosity by means of the empirical relation given by
\citet{Rosa-Gonzalez+02}.
For the galactic disk, the SFR measured by \citet{Leroy+08} has been used.

Electron densities ($n_{\rm e}$) of the emitting gas in the CNSFRs have been
computed from the [S{\sc ii}]$\lambda\lambda$6717/6731\,\AA\ line ratio, and 
electron temperatures refer to $T_{\rm e}$([O{\sc iii}]) \citep{Diaz+07}.
For the eighth region, values of 400~cm$^{-3}$ and 5000~K have been assumed.
The ambient ISM densities of the central diffuse region and the disk are 10
    and 1~cm$^{-3}$, respectively, with a temperature of $10^4$~K.
The associated filling factors ($f$) are given by
expression~(\ref{eqFillFactor}). 
As explained in Appendix~\ref{secThermal}, the predicted optical depth
($\tau_{\rm 1.42\,GHz}^{\rm brems}$) and flux density ($S_{\rm 1.42\,GHz}^{\rm
  brems}$) at 1.42~GHz arising from thermal bremsstrahlung take into account
self-absorption, although the effect is found to be mild at this frequency. 

Star formation rates have been estimated from the \Ha luminosity
\citep{Rosa-Gonzalez+02}, and the expected number of SNRs 
($N_{\rm SNR}$) has been computed according to equation~(\ref{eqNumberSNR}),
assuming an ambient ISM density of 10~cm$^{-3}$ for the CNSFRs.
The number of SNRs brighter than 0.3~mJy ($N_{\rm SNR}^{( 0.3~{\rm mJy} )}$) is
given by expression~(\ref{eqN_SNR_thr}), and the total synchrotron emission
from the whole population of unresolved supernovae ($S_{\rm 1.42\,GHz}^{\rm
  synchr}$) has been estimated from the empirical relation between the star
formation rate and the 1.4~GHz luminosity density reported by \cite{Yun+01}.

As shown in the previous section, our analysis of the sub-arcsecond spatial
resolution radio map of NGC\,3351 shows the presence of 23 sources in the
L-band. 
Three of them (R1, R2 and R3) are located within the central 100~arcsec in
radius. 
They have clear counterparts in the FIRST survey \citep{White+97} and the VLA
data analysed by \citet{Paladino06}, but only R1 seems to be visible at other
wavelengths. It is unresolved by MERLIN, putting an upper limit of $\sim10$ pc
on its size, but its variability (Section~\ref{secResults}) suggests that it
is $<3$ pc.  As shown in Figure~\ref{R2+R3-multiw}, neither R2 nor R3 display
significant emission in X-ray, UV, optical, or IR images.  These sources are
significantly extended (see Figure~\ref{extended-sources}).
They are separated by about half an arcmin and could possibly be associated
with jets from a background AGN. 
FIRST contains about 7 sources deg$^{-2}$ brighter than 20 mJy; the
probability of finding one of them within a region of 100~arcsec radius chosen
at random is only $\sim 2$ per cent, but their observed fluxes and extents are
difficult to reconcile with a stellar origin. 
Moreover, the chance of two unrelated sources of such high brightness
occurring so close together is extremely low, favouring the background AGN
scenario. Thus, the only radio point source above 0.3~mJy that may be related
to star formation in NGC\,3351 is R1. 

In particular, we have not detected any point source in the nuclear region.
The thermal emission predicted for each of the CNSFR complexes ranges from
0.03 to 0.36\,mJy, slightly below our detection limit\footnote{For the
requested observing time of 48 hours, the expected noise level would have
been $\sigma_{\rm rms}=0.03$~mJy under good weather conditions.}. The estimated
fluxes are fairly robust with respect to the adopted sizes. Smaller diameters,
appropriate for the individual knots \citep[see e.g.][]{Hagele+07}, yield
higher filling factors and optical depths, smaller areas, and, then, similar
fluxes. On the other hand, a number of supernovae is expected to be present in
these regions. 
The expected probability of finding a SNR brighter than 0.3\,mJy in the
circumnuclear ring is relatively high ($\sim 50$~per cent), and the synchrotron
flux associated to the unresolved population within each CNSFR (last column of
Table~\ref{tab:regions}) would have been easily detected by MERLIN. 
Therefore, the observed lack of emission suggests that these systems are
younger than the lifetimes of the most massive stars \citep[see
e.g.][]{Rosa-Gonzalez+07a}. This seems also to be the case in similar CNSFRs
observed in other galaxies \cite[e.g.][in NGC\,6951]{Dors+08}.

Regarding the star formation activity in the rest of the central region, the
predicted number of bright SNR is also $\sim 0.5$, and the total synchrotron
flux is estimated to be about 36~mJy. 
Indeed, VLA observations at a minimum resolution of $4.5 \times 4.1$~arcsec
reveal a large central core, about 20~arcsec in size, featuring a double
emission peak with maximum surface brightness about 10$^{10}$ mJy sr$^{-1}$
\citep{Paladino06}, $\sim 0.05$ mJy per MERLIN beam (close to the rms noise of
our observations) if the emission was uniformly distributed.
If star formation takes place in compact knots, similar in size to those in
the CNSFRs, this emission ought to display conspicuous small-scale structure
that should be detected by MERLIN.
However, MERLIN is only sensitive to scales smaller than $\sim2$ arcsec, which
is $\sim 100$ pc at the distance of NGC\,3351.
Therefore, the relativistic electrons generated by the supernova explosions
must have diffused over distances longer than our maximum resolution radius of
$\sim 50$~pc, and the true diffusion scale 
would thus be somewhere in between this value and the 300~pc (6~arcsec) where
the radio-IR correlation is observed by \citet{Paladino06}. 

The total flux density at 1.42\,GHz, integrated over the whole galaxy, is
predicted to be $\sim 0.14$~Jy. According to equation~(\ref{eqN_SNR_thr}), the
expected number of radio sources above 0.3\,mJy in the disk of NGC\,3351 is of
the order of one, consistent with the interpretation that R1 is a young
SNR.
However, an increase in flux of a factor of two between the FIRST
observations and our data seems difficult to justify in this scenario.
Two possible explanations would be that FIRST caught the supernova near the
time of the explosion, where the radio continuum was still rising, and we are
now observing a very young remnant, as suggested by its high luminosity.
An alternative possibility would be that the radio flux has increased due to
density variations in the ambient medium.
Some Galactic nebulae, such as Cas A \citep{Erickson+75,Read+77,Chevalier+78}
and GK Per \citep{Anupama+05} also show anomalous
increases in flux density, albeit at lower frequencies.
Ideally SNRs could be identified based on the radio spectral index, and
future observations will be sought based on the results of the current work.

\section{Summary}
\label{secConclus}

We present high spatial resolution radio continuum observations
at 1.42\,GHz acquired with the Multi-Element Radio Linked Interferometer
Network (MERLIN) in and around NGC\,3351.
We found 23 radio sources in a field of 35 by 22\,arcmin centred in this
galaxy. 6 of them are projected within its D$_{25}$ isophote, and 3 are
located inside the central 100\,arcsec in radius. 
Two of these three are significantly extended, and the other one is the only
source with a previously detected counterpart at other wavelengths.
Our results suggest that this radio source might correspond to a young
supernova remnant, while the other two are probably related to jets from a
background AGN.

We do not detect individual supernovae or SNRs in the central region of
the galaxy.
The CNSFRs are too young (less than a few Myr) to host supernovae,
and the diffusion length of the relativistic electrons in the ISM seems to be
larger than our maximum resolution of 50 pc.
Detecting the thermal bremsstrahlung emission from the circumnuclear \HII
regions requires deeper observations.
\emph{e}-MERLIN will allow NGC\,3351 to be mapped at a sensitivity of
$\sigma_{\mathrm{rms}} \sim 10 \mu$Jy beam$^{-1}$ in a similar observation,
thanks not only to the increased sensitivity provided by 0.5 MHz bandwidth
around 1.5 GHz but also to the new correlator which will allow the entire
primary beam to be imaged without distortion.

\section*{Acknowledgments}

We are grateful to an anonymous referee for his/her constructive
comments and revision of our manuscript.

MERLIN is a National Facility operated by the University of Manchester at
Jodrell Bank Observatory on behalf of STFC.

Some of the data presented in this paper were obtained from the Multimission
Archive at the Space Telescope Science Institute (MAST) and the Hubble Legacy
Archive, which is a collaboration between the Space Telescope Science
Institute (STScI/NASA), the Space Telescope European Coordinating Facility
(ST-ECF/ESA) and the Canadian Astronomy Data Centre (CADC/NRC/CSA). STScI is
operated by the Association of Universities for Research in Astronomy, Inc.,
under NASA contract NAS5-26555. Support for MAST for non-HST data is provided
by the NASA Office of Space Science via grant NAG5-7584 and by other grants
and contracts.

This research has made use of data obtained from the Chandra Data Archive, and
software provided by the Chandra X-ray Center (CXC) in the application package
CIAO.

This research has made use of the NASA/IPAC Extragalactic Database (NED) which
is operated by the Jet Propulsion Laboratory, California Institute of
Technology, under contract with the National Aeronautics and Space
Administration and of the SIMBAD database, operated at CDS,
Strasbourg, France. 
HST

Financial support for this work has been provided by the Spanish
\emph{Ministerio de Educaci\'on y Ciencia} (AYA2007-67965-C03-03). Partial
support from the Comunidad de Madrid under grant S-0505/ESP/000237 (ASTROCAM)
is acknowledged.

 \bibliographystyle{mn2e}
 \bibliography{references}

\appendix

\section{Thermal bremsstrahlung from \HII regions}
\label{secThermal}

Thermal bremsstrahlung from \HII regions can be a potential signature of star formation in the radio band.
The optical depth through an ionised hydrogen cloud can be approximated by \citep{MezgerHenderson67}
\begin{equation}
 \tau_\nu \approx 3.3 \times 10^{-7}
\left( \frac{\nu}{\rm GHz} \right)^{\!\!-2.1}
\left( \frac{T_{\rm e}}{10^4\rm K} \right)^{\!\!-1.35}
\left( \frac{EM}{\rm cm^{-6}~pc} \right)
\end{equation}
where the emission measure is defined as the integral along the line of sight
\begin{equation}
EM = \int \Ne n_{\rm i}\ \dd l \approx \int \Ne^2\ \dd l
\end{equation}
with $\Ne$ and $n_{\rm i}$ denoting electron and ion density, respectively.
For a pure hydrogen, fully ionised plasma, $\Ne=n_{\rm i}$.
Including helium would introduce a correction of the order of 10 per cent, and the contribution of heavier elements is negligible.

The overall extent of the CNSFRs in NGC\,3351 is $\sim 100$~pc, but most of the emission comes from a few compact knots of $(1-5)$~pc radius.
The characteristic electron density of the emitting gas is accurately measured by the ratio between the optical sulphur lines [S{\sc ii}]$\lambda\lambda$6717/6731~\AA, with typical values of $\sim400$~cm$^{-3}$ \citep{Diaz+07}.
The volume filling factor $f$ of the (unresolved) knots can be derived from \Ha luminosity of the entire region
\begin{equation}
 L(\Ha) = E_\alpha \sigma_\alpha \Ne^2 f \frac{\pi}{4} D^3
\end{equation}
where $E_\alpha = 3\times10^{-12}$~erg is the energy of the \Ha transition,
\begin{equation}
\frac{ \sigma_\alpha }{ \rm cm^3\ s^{-1} } = 1.17 \times 10^{-13}
\left( \frac{T_{\rm e}}{10^4\rm K} \right)^{\!\!-0.87}
\end{equation}
is the rate coefficient, and a cylindrical geometry with both diameter and depth $D$ has been assumed for simplicity.
The uncertainty associated with this assumption is high (e.g. the difference
with respect to a perfect sphere of diameter $D$ is a factor of $1.5$), and it
constitutes an important source of error in the following estimates. 
With this caveat in mind, the filling factor of the emitting gas is approximately given by the expression
\begin{equation}
f \approx 1.2
\left(\!\frac{ L(\Ha)  }{\rm 10^{31}\,erg\,s^{-1}} \right)
\left(  \frac{   \Ne   }{\rm cm^{-3}            } \right)^{\!\!-2}
\left(\!\frac{    D    }{\rm pc                 } \right)^{\!\!-3}\!
\left(\!\frac{T_{\rm e}}{\rm 10^4\,K             } \right)^{\!0.87}
\label{eqFillFactor}
\end{equation}
If there is a small number $N$ of identical knots within the region, so that they do not overlap along the line of sight, the total emission measure would be
\begin{equation}
 EM = \Ne^2 \left( \frac{f}{N} \right)^{\!\!1/3} D
\end{equation}
and the total area that they subtend on the sky is
\begin{equation}
 A = N \frac{\pi}{4} \left( \frac{f}{N} \right)^{\!\!2/3} D^2
\end{equation}
whereas for large $N$, assuming that the knots are randomly distributed,
\begin{equation}
 EM = \Ne^2 f D
\end{equation}
and
\begin{equation}
 A = \frac{\pi}{4} D^2
\end{equation}
The two cases $N=1$ and $N\to\infty$ can be considered as reasonable upper and lower limits to the emission measure, respectively.
The radio optical depth is thus bracketed by
\begin{eqnarray}
\nonumber
\tau_\nu &\approx& 3.3 \times 10^{-7}\,
\left( \frac{ \nu }{ \rm GHz } \right)^{\!\!-2.1}
\left( \frac{ T_{\rm e} }{ \rm 10^4\,K } \right)^{\!\!-1.35}
\\ &&\times
\left( \frac{ \Ne }{ \rm cm^{-3} } \right)^{\!2}
\left( \frac{ D }{ \rm pc } \right)
f^\eta
\label{eqTauFF}
\end{eqnarray}
with $1/3<\eta<1$.
Taking the geometric average, $\eta=2/3$, the relative error incurred in the optical depth is smaller than a factor $f^{1/3}$.

The radio-continuum surface brightness is given by the expression
\begin{equation}
\Sigma_\nu = B_\nu\, (1 - e^{-\tau_\nu})
\end{equation}
where $B_\nu\approx\frac{2kT\nu^2}{c^2}$ is the Planck function for a black body at temperature $T$ in the Rayleigh-Jeans approximation, and the flux density at the Earth is
\begin{eqnarray}
\nonumber
\frac{ S_\nu }{\rm Jy}\!\!\!\!&=&\!\!\!\! \frac{ A\, \Sigma_\nu }{ d^2 }
\\ &\approx&\!\!\!\!
2.4\times 10^8\!
\left( \frac{ \nu }{ \rm GHz } \right)^{\!\!2}\!
\left(\!\frac{ T_{\rm e} }{\rm 10^4\,K } \right)\!\!
\left( \frac{D}{d} \right)^{\!\!2} f^{1-\eta}
\, (1 - e^{-\tau_\nu})
\label{eqThermalFlux}
\end{eqnarray}
In the optically thin case, $\Sigma_\nu \approx B_\nu \tau_\nu$ implies
\begin{equation}
\frac{ S_\nu }{\rm Jy} \approx 79
\left(\! \frac{ \nu }{ \rm GHz } \right)^{\!\!-0.1}
\!\!\left(\!\! \frac{ T_{\rm e} }{ \rm 10^4\,K }\!\! \right)^{\!\!-0.35}
\!\!\left(\! \frac{D}{d} \right)^{\!\!2}
\!f
\left(\! \frac{ \Ne }{ \rm cm^{-3} } \!\right)^{\!\!2}
\!\left(\! \frac{ D }{ \rm pc } \right)
\end{equation}
and one recovers the well-known relation
\begin{equation}
\frac{ S_\nu }{\rm Jy} \approx 1.1 \times 10^9
\left( \frac{ \nu }{ \rm GHz } \right)^{\!\!-0.1}
\!\left(\! \frac{ T_{\rm e} }{ \rm 10^4\,K } \!\right)^{\!\!0.52}\!
\left(\! \frac{ F(\Ha) }{ \rm erg\,s^{-1}\,cm^{-2} } \!\right)
\end{equation}
between the radio continuum and \Ha fluxes.
In the general case, the radio-continuum flux density may be estimated by combining expressions~(\ref{eqFillFactor}), ~(\ref{eqTauFF}), and~(\ref{eqThermalFlux}).

\section{Synchrotron emission from supernova remnants}

Assuming that all stars more massive than $M_{\rm SN}\simeq 8~\Msun$ explode
as {\bf core collapse (Type~II)} supernovae, the number of such events per
unit time is determined 
by the SFR $\dot M_*$ and the initial mass function (IMF) of the stellar
population, 
\begin{equation}
\nu_{\rm SN}(t) =
\frac{ \int_{M_{\rm SN}}^{M_{\rm max}} \dot M_*(t-\tau_*(m))\ \psi(m)\ \dd m }
     { \int_{M_{\rm min}}^{M_{\rm max}} m\ \psi(m)\ \dd m }
\end{equation}
where $\tau_*(m)$ is the lifetime of a star of mass $m$.
For a \cite{Salpeter55} IMF, $\psi(m)\propto m^{-2.35}$, with minimum and
maximum stellar masses $M_{\rm min}\sim0.1~\Msun$ and $M_{\rm
  max}\sim100~\Msun$, respectively. 
If the SFR is constant over the period $\tau_*(M_{\rm SN})\sim40$~Myr that the least massive supernovae take to explode,
\begin{equation}
\nu_{\rm SN}(t) \approx
\frac{ \dot M_*(t) \int_{M_{\rm SN}}^{M_{\rm max}} \psi(m)\ \dd m }
     { \int_{M_{\rm min}}^{M_{\rm max}} m\ \psi(m)\ \dd m }
\approx 7.4\times10^{-3} \frac{\dot M_*(t)}{\Msun}
\label{eqNuSN}
\end{equation}

SNRs can be detected as compact synchrotron sources during the characteristic time
\begin{equation}
\frac{ \tau_{\rm SNR} }{ \rm yr } \approx
3.4\times10^4
\left( \frac{ E_{\rm SN} }{ 10^{51}~{\rm erg} } \right)^{4/17}
\left( \frac{ n }{ \rm cm^{-3} } \right)^{-9/17}
\label{eqTauSNR}
\end{equation}
of the adiabatic expansion phase, where $E_{\rm SN}$ denotes the energy input by the original supernova explosion, and $n$ is the density of the ambient gas \citep{Woltjer72}.
The total number of SNR observable at a given time in the radio band would be
\begin{eqnarray}
\nonumber
N_{\rm SNR}(t) \!\!\!&=& \int_{t-\tau_{\rm SNR}}^{t}\!\! \nu_{\rm SN}(t')\ \dd t'
\ \approx\ \nu_{\rm SN}(t)\, \tau_{\rm SNR}\\
&\approx&
\!\!\!250\, \frac{\dot M_*(t) }{\rm \Msun~yr^{-1} }
\left( \frac{ n }{ \rm cm^{-3} } \right)^{\!\!-9/17}
\label{eqNumberSNR}
\end{eqnarray}
assuming $E_{\rm SN} = 10^{51}$~erg.
The approximation of constant SFR for the last 40~Myr and a density
$n\sim1$~cm$^{-3}$ are probably valid for most regions of the galaxy disk,
where a moderate star formation surface density of the order of $\dot\Sigma_*
\sim (1-3)\times10^{-3}~{\rm \Msun~yr^{-1}~kpc^{-2} }$ is observed without
prominent signatures of intense starbursts \citep[see e.g.][]{Leroy+08}. 
The expected supernova surface density throughout the disk is thus
\begin{equation}
\Sigma_{\rm SNR} \approx 250\ \frac{\dot\Sigma_*}{\rm \Msun~yr^{-1} }
\approx 0.5\ {\rm kpc^{-2}}
\end{equation}
and the total number of SNR in a corona between 0.6 and 10~kpc (12 and 200
arcsec) from the galactic centre would be about 125, corresponding to an
integrated SFR of~$0.5~{\rm \Msun~yr^{-1}}$. 

The inner 600 pc of NGC\,3351 are also forming stars at a rate of
$\sim0.5~{\rm \Msun~yr^{-1}}$. 
A large fraction \citep[about $0.24~{\rm \Msun~yr^{-1}}$, see][]{Hagele+07}
can be accounted for by the young starbursts of the CNSFRs, yielding a diffuse
contribution of~$\sim0.26~{\rm \Msun~yr^{-1}}$. 
Although equation~(\ref{eqNumberSNR}) predicts a similar number of supernovae
in the central region than in the whole disk, the density of the ISM is
arguably higher. 
Using $n=10$~cm$^{-3}$, the expected number of SNR becomes $\sim 40$.
Concerning the individual CNSFRs, the observed values of $\dot M_* \sim
(0.01-0.06)~{\rm \Msun~yr^{-1}}$ suggest that most bursts should contain SNRs.
However, one has to bear in mind that the star formation activity may have
varied considerably over the last $40$~Myr. 
In the most extreme case, there is the possibility that all bursts are younger
than 3~Myr, and therefore no supernova could have exploded yet. 
On the other hand, it is likely that the gas inside and around the CNSFRs is
considerably denser than 10~cm$^{-3}$, leading to a shorter radio lifetime
$\tau_{\rm SNR}$ and a smaller number of radio-emitting SNR. 

Regardless of the number of sources actually present in the galaxy, they must be bright enough in order to be included in our sample.
A relation between the radio surface brightness $\Sigma_\nu$ and the diameter $D$ of SNRs has long been established both theoretically \citep{Shklovskii60} and observationally \citep{ClarkCaswell76}.
At 1~GHz, \citet{Urosevic+05} obtain
\begin{equation}
\frac{ \Sigma_{\rm 1\,GHz} }{\rm W~Hz^{-1}~m^{-2}~sr^{-1} } =
8.84 \times 10^{-16} \left( \frac{ D }{ \rm pc } \right)^{\!\!-3.2}
\end{equation}
combining data from different nearby galaxies.
To estimate the flux density of an extended source, one has to multiply its surface brightness by the solid angle subtended by the MERLIN beam,
\begin{equation}
\frac{ S_{\rm 1.42\,GHz} }{\rm mJy~beam^{-1}}
\approx \pi \left( \frac{ \theta_{\rm FWHM} }{ 2 } \right)^{\!\!2} \Sigma_{\rm 1.4\,GHz}
= 55 \left( \frac{ D }{ \rm pc } \right)^{\!\!-3.2}
\end{equation}
where we have used $\theta_{\rm FWHM}=0.2$~arcsec and a spectral index $\alpha=0.5$ to convert 1~GHz fluxes to $1.42$~GHz.
At the adopted distance of NGC\,3351, an angle of 0.2~arcsec corresponds to a physical size of $\sim 10$~pc and a flux density of $35~\mu$Jy~beam$^{-1}$, below the noise level of the present data.
Therefore, any SNR above our detection threshold must appear as an unresolved point source with flux
\begin{equation}
\frac{ S_{\rm 1.42\,GHz} }{\rm \mu Jy}
\approx \pi \left( \frac{ D/2 }{ d } \right)^2 \Sigma_{\rm 1.4\,GHz}
= 577\, \left( \frac{ D }{ \rm pc } \right)^{\!-1.2}
\end{equation}
For a source as bright as R1 (4~mJy), the expected diameter is only $0.20$~pc, for which the SNR would still be in the free expansion phase.
Assuming an average expansion velocity of $\sim3000$~km~s$^{-1}$, the age of this system should be about 35 years.
Substituting this number instead of $\tau_{\rm SNR}$ in equation~(\ref{eqNumberSNR}), one obtains
\begin{equation}
N_{\rm SNR}( S_{\rm 1.42\,GHz} \!\ge\!4~{ \rm mJy } )
\,\approx\, 0.25\, \frac{ \dot M_*(t) }{ \rm \Msun~yr^{-1} }
\end{equation}
Lowering the threshold to $300~\mu$Jy, the diameter increases to 1.72~pc, and the expected number of sources becomes
\begin{equation}
N_{\rm SNR}( S_{\rm 1.42\,GHz} \!\ge\!0.3~{ \rm mJy } )
\,\approx\, 2\, \frac{ \dot M_*(t) }{ \rm \Msun~yr^{-1} }
\label{eqN_SNR_thr}
\end{equation}

\end{document}